\documentclass[twocolumn,showpacs,preprintnumbers,amsmath,aps]{revtex4}
\usepackage{epsfig}
\begin{document}

\title{On effects of regular S=1 dilution of S=1/2
  antiferromagnetic Heisenberg chains by a quantum Monte Carlo simulation}

\author{
Fengping Jin, Zhaoxin Xu\footnote{E-mail: zxxu@zimp.zju.edu.cn.}, Heping Ying and Bo Zheng
}

\affiliation{
Zhejiang Institute of Modern Physics, Zhejiang University,
Hangzhou 310027, P.R.China \\
}


\date{\today}

\begin{abstract}
The effects of regular S=1 dilution of S=1/2 isotropic
antiferromagnetic chain are investigated by the quantum Monte Carlo
loop/cluster algorithm. Our numerical results show that there are
two kinds of ground-state phases which alternate with the
variation of $S^1=1$ concentration. When the effective spin of a unit
cell is half-integer, the ground state is ferrimagnetic with
gapless energy spectrum and the magnetism becomes weaker with
decreasing of the $S^1$ concentration $\rho = 1/M$. While it is
integer, a non-magnetic ground state with gaped spectrum emerges
and the gap gradually becomes narrowed as fitted by a relation of
$\Delta \approx 1.25\sqrt{\rho}$.
\end{abstract}
\maketitle
\section{Introduction}
The effects of substitutions of magnetic impurities on the
antiferromagnetic spin chain have attracted great interests in the
past decade. It has been shown theoretically that the ground state
properties vary with different dilution cases. For the random
substitutions, the most interested case is of the S=1/2 impurities in
Haldane chain \cite{Sorensen}. For example,  inelastic neutron scattering
experiment on the compound Y$_2$BaNiO$_5$ substituting Ca$^{2+}$
for Ni$^{2+}$ \cite{Ditusa} show a substantial increase of the
spectral function below the Haldane gap to indicate the
creation of states below the energy of the spin gap. This effects
are also studied by numerical works by S. Wessel \cite{Wessel}.
For regular substitutions, these systems are the mixed-spin chains
which have been extensively studied by many authors in the past a
few years. Analytical methods of non-linear sigma model, mean field
theory and spin-wave method \cite{Yamamoto2,Fukui,Takano,Wu} as well as
numerical works by density matrix renormalization group
\cite{Pati} and quantum Monte Carlo \cite{Tonegawa, Yamamoto} have been
applied extensively for such systems. So far, it is well known that the
topology of spin arrangements in the mixed chains plays an
essential role on the ground state properties and thermodynamics
in the mixed-spin systems. 

Experimently, many Quasi-1D mixed-spin materials have
been synthesized in the past two decades, such as ACu(pba)(H$_2$
O)$_3$ $\cdot$ n(H$_2$ O) and ACu(pbaOH)(H$_2$ O)$_3$ $\cdot$
nH$_2$O (where
pba=1,3-propylenebis(oxamato),pbaOH=2-hydroxo-1,3-propylenebis and
A= Ni,Fe,Co,Mn,Zn). These materials contains two different
transition metal ions per unit cell, and their properties were
studied as ferrimagnetic chains
\cite{Kahn1,Kahn2,Verdaguer,Gleizes,Pei,Hagiwara2}.  The
experiment results imply that the magnetic properties of the mixed-spin
compounds can all be described by a Heisenberg model with
nearest-neighbor antiferromagnetic coupling as
\begin{equation}
H=\sum_{i=1}^{N} J_i S_i \cdot S_{i+1},
\label{Hamiltonian}
\end{equation}
where $S_i$ denotes a spin-$S$ moment at site $i$, $N$ is system size and $J_i>0$.
T. Fukui and N. Kawakami \cite{Fukui} have studied spin chain composed
by a periodic array of impurities $S^1$ embedded in the host $S^2 \neq
S ^1$ spin chain with the period $M$, i.e.,
\begin{equation}
\underbrace{S^1 \otimes S^2 \otimes S^2 \otimes \ldots \otimes
S^2}_{\mbox{M}} \otimes S^1 \otimes S^2 \otimes \ldots \otimes
S^2.
\label{spin_arrange}
\end{equation}

The dilutions of the model denoted by the impurity concentration
$\rho$ has two limits: ({\it i}) $\rho=0$, the undoped pure
antiferromagnetic $S^2$ chain, it has a non-magnetic ground state;
({\it ii}) $\rho=0.5$, the alternating spin chain of $S^1$ and
$S^2$. According to Marshall theorem and Lieb-Schultz-Mattis (LSM)
theorem \cite{Lieb}, ground state of the doped cases are specified
by the spin quantum number $S=0(|S_1-S_2|N/M)$ for $M=odd$ or
$even$, it is either a spin singlet or ferrimagnetic. If the
effective spin in a unit composed of $M$ spins $S_{eff}$ is
half-integer, so the system has a gapless energy spectrum. But
when $S_{eff}$ is integer, LSM theorem fails to predict the energy
spectrum to be gaped or gapless. By applying non-linear $\sigma$
model, it is found that the system has an energy gap when the
$S_{eff}$ is integer\cite{Fukui, Takano}. But details of ground
state properties and thermodynamics can not be given by
 non-linear $\sigma$ model analyses. 

The authors of present paper
have recently studied the model (1) with the case of $S^1=1/2$ and
$S^2=1$ by applying quantum Monte Carlo simulations \cite{Wang},
where the numerical results reveal
different non-trivial magnetic properties happened between two
kinds of diluting cases, i.e. for $odd$ $S^2=1$ spins in a unit, system
has magnetic ground state and it shows ferrimagnetic features;
while for $even$ $S^2=1$ spins in a unit, systems behave non-magnetic
ground states with antiferromagnetic-like features. For both the
$odd-even$ cases, the ground states are {\it gapless} steadily. And the system gradually
transits from the ferrimagnetic ground state of the alternating
$S^1$-$S^2$ chain to the disordered ground state of pure $S=1$ chain in two
different tendencies. In this Letter, we study an opposite case with $S^1=1$
and $S^2=1/2$. Previous analytical work predicted that if $odd$
$S^2=1/2$ spins in a unit, the effective spin $S_{eff}$ is
half-integer, the ground state is ferrimagnetic with a gapless
energy spectrum; while if $even$ $S^2=1/2$ in a unit, $S_{eff}$ is
integer, the ground state is non-magnetic and the system has an energy gap.
Our numerical study will focus on how the ground state properties
depend on the concentration and the finite temperature magnetic properties
evolute with decreasing of the $S^1=1$ concentration $\rho$.

\section{Calculation and Results}
We use the efficient continuous imaginary time version of loop cluster
algorithm to perform the quantum Monte Carlo simulation
\cite{Beard}, which has been successfully applied for the other
mixed-spin chains \cite{Xu, Wang}.
We confine our calculation to isotropic antiferromagnetic coupling
cases, i.e. $J_i = J >0$ in equation (\ref{Hamiltonian}), and the positions of
spin $S_1=\frac{1}{2}$ and $S_2=1$ are arranged as represented in
equation (\ref{spin_arrange}) with $M$ taking the values from 2 to
11. We carry out $10^5$ Monte Carlo
steps for measuring physical quantities after $10^3$ Monte Carlo steps
for the thermalization. In order to clearly explore the ground state
properties, the simulations are performed at the very low temperature
$\beta = 1/T= 200$ for system sizes $L>200$ in condition of even number of
unit. The physical quantities we measure are the ground state energy $E_G$, the uniform magnetic
susceptibility $\chi_u$ and staggered susceptibility $\chi_s$ by
using the improved estimators in the loop cluster algorithm, e.g. ,
\begin{equation}
<\chi>=\frac{\beta}{4V} \Bigl \langle \sum_{cluster \\ c} w_t(c)^2 \Bigr \rangle_{MC},
\end{equation}
\begin{equation}
<\chi_s>=\frac{1}{4V\beta} \Bigl \langle \sum_{cluster \\ c} |C|^2 \Bigr \rangle_{MC},
\end{equation}
where $w_t(c)$ is winding number of cluster $c$, and $|C|$ is the cluster size.
The magnetization and staggered magnetization are estimated by
\begin{equation}
<M^2>= \Bigl \langle 3(\sum_i S_i^z)^2 \Bigr \rangle_{MC}
\end{equation}
and
\begin{equation}
<M_s^2>=\Bigl \langle 3(\sum_i(-1)^i S_i^z)^2 \Bigr \rangle_{MC}.
\end{equation}
The energy gap $\Delta$ is also estimated in the way given by Todo, \cite{Todo}
\begin{equation}
\bigtriangleup = \lim_{L \rightarrow \infty} \frac{1}{\xi_{\tau,0}(L)},
\end{equation}
where $\xi_{\tau,0}$ is the correlation length in the imaginary time direction.

\begin{figure}[ht]
\vspace{0.4cm}
\epsfxsize=6cm\epsfysize=4cm \centerline{\epsffile{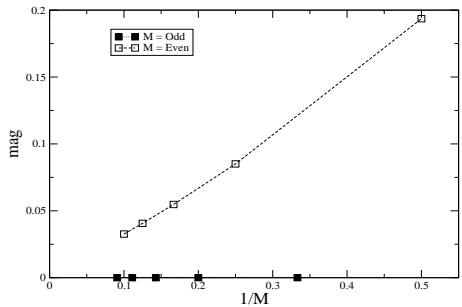}}
\caption{The magnetization versus $\rho=1/M$,
  i.e. the diluting concentration of $S^1$. The filled squares present
  the cases of $M=even$ and the empty squares for $M=odd$. }
\label{mz}
\end{figure}

\begin{figure}[ht]
\vspace{0.4cm}
\epsfxsize=6cm\epsfysize=4cm \centerline{\epsffile{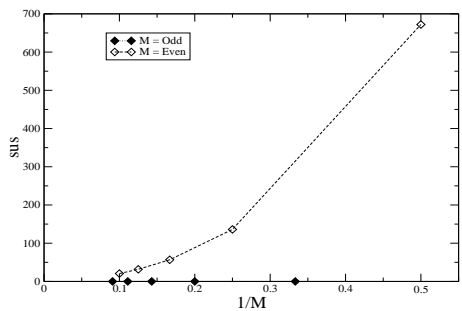}}
\caption{The magnetic susceptibility versus
$\rho=1/M$. The filled diamonds
  are for the cases of $M=even$ and the empty diamonds for $M=odd$.}
\label{sus_u}
\end{figure}

The results for magnetizations and uniform susceptibility are plotted in Fig. \ref{mz}
and Fig. \ref{sus_u}. We find that the magnetic properties are apparently
different for two cases of $M=odd$ and $M=even$.  When $M=even$, the magnetization
is finite and approaches zero linearly with decreasing of
$\rho$. While $M=odd$, the magnetization remains almost at zero
value. On the other hand, it can be observed from our results that the
uniform susceptibilities $\chi_u$ is finite for $M=even$, but it vanishes when $M=odd$.
Thus there is magnetic long-range order (LRO) in the ground state when
$M=even$, but the order is absent when $M=even$.

\begin{figure}[ht]
\vspace{0.4cm}
\epsfxsize=6cm\epsfysize=4cm \centerline{\epsffile{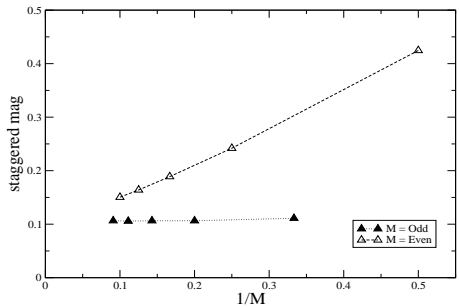}}
\caption{The staggered magnetization versus $\rho=1/M$. The filled
  triangles are for the cases of $M=even$ and empty ones for $M=odd$.}
\label{smz}
\end{figure}

\begin{figure}[ht]
\vspace{0.4cm}
\epsfxsize=6cm\epsfysize=4cm \centerline{\epsffile{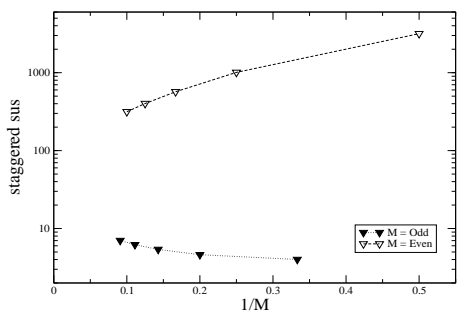}}
\caption{The staggered susceptibility versus $\rho=1/M$. The filled
  triangles are for the cases of $M=even$ and empty ones for $M=odd$.}
\label{sus_s}
\end{figure}

We further estimate the staggered magnetization and its susceptibility
as a function of concentration shown in Fig. \ref{smz} and
Fig. \ref{sus_s} respectively.
The two observables are both finite for $M=odd$ and $M=even$ cases,
but the data for the cases of $M=even$ have much stronger values than
the cases of $M=odd$.

\begin{figure}[ht]
\vspace{0.4cm}
\epsfxsize=6cm\epsfysize=4cm
\centerline{\epsffile{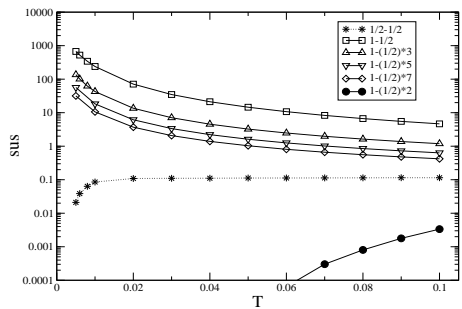}}
\caption{The uniform magnetic susceptibility $\chi_u$ versus
  temperature. The empty symbols are for the cases of $M=even$ and
  filled ones for $M=odd$. The star present the result of undoped $S=1/2$
  antiferromagnetic Heisenberg chain.}
\label{sus_T}
\end{figure}

In order to confirm the results observed above, we begin to
investigate the finite temperature uniform magnetic
susceptibility. As displayed in Fig. \ref{sus_T},
one can easily find that $\chi_u$ diverges when the temperature
$T=1/\beta$ goes to zero in the cases $M=even$. This is the
typical behavior of a system with magnetic LRO. In the cases
$M=odd$, all the $\chi_u$ approach zero when $T \rightarrow 0$, a
remarkable evidence to reveal the existence of the energy gap.

Up to now, our results verify numerically that there are magnetic
LRO and antiferromagnetic LRO in the ground states when $M=even$.
They clearly show that the ground states are ferrimagnetic in such
cases. While for $M=odd$, there should exist of spin liquid phases
denoted by the vanish of the magnetizations. Consequently we
believe our numerical results consist correctly with the previous
analytical predictions. More important, one can easily see that
the magnetism decreases with decreasing of impurity concentration
in the case of $M=even$. But there is not notable change of the
magnetic properties when the $S^1=1$ concentration decreases as
$M=odd$.

\begin{figure}[ht]
\vspace{0.3cm}
\epsfxsize=6cm\epsfysize=4cm \centerline{\epsffile{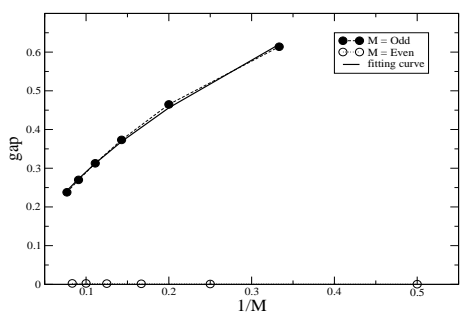}}
\vspace{0.3cm}
\caption{The energy gap versus $\alpha$, i.e. the diluting concentration of
$S^1$. The filled circles present the cases of $M=odd$ and empty ones
present the cases of $M=even$, the dashed lines are guide for eyes. The
thin line is the fitting curve $\Delta=1.25\sqrt{\rho}-0.1$, the
correlation coefficient of this curve to the empty circles is $0.999$.}
\label{gap}
\end{figure}

Next, we consider the feature of the energy gap $\Delta$ on different regular
dilutions. Not surperised for us, the energy gap is closed when $M$ is
$even$ and it opens again while $M$ is $odd$ as shown in
Fig. \ref{gap}. These results is consist with the
prediction by non-linear $\sigma$ model and LSM theorem \cite{Fukui,
  Takano}.
It is interesting that the energy gap $\Delta$ tends to be narrow
as decreasing of $S^1=1$ concentration when $M=odd$. We confirm
such behavior by fitting $\Delta$ to the curve of
$1.25\sqrt{\rho}$ as one can see in Fig. \ref{gap}.

\begin{figure}[ht]
\vspace{0.4cm}
\epsfxsize=6cm\epsfysize=4cm \centerline{\epsffile{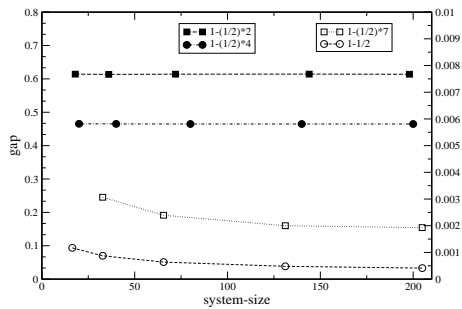}}
\caption{The finite-size effect of energy gaps. The two upper lines of
filled symbols are for cases of $M=odd$ whose values are presented by
the y-axis at left-hand side, and the two lower lines of
empty symbols are for $M=even$ whose values are presented
right-hand side y-axis.}
\label{size}
\end{figure}

Moreover, we show the finite-size effect of $\Delta$ results for
several cases with $L$ increasing in Fig. \ref{size}. In our
estimations, although the gaps are not exact closed for $M=even$
due to the finite-size simulations, we find the data of the gaps
decrease fast than $L^{-1}$, so it is obvious that
the gaps will trend to zero as $L \rightarrow \infty$. For the
cases $M=odd$, where the gap opens all the time, there is almost no
finite-size effect. 

In order to identify the ground state phases, we calculate the
valence-bond-solid (VBS) \cite{Affleck} order parameter
\begin{equation}
z\equiv \langle \exp[i\frac{2\pi}{N}\sum^N_{j=1}jS^z_j]\rangle,
\end{equation}
\begin{figure}[ht]
\vspace{0.4cm}
\epsfxsize=6cm\epsfysize=4cm \centerline{\epsffile{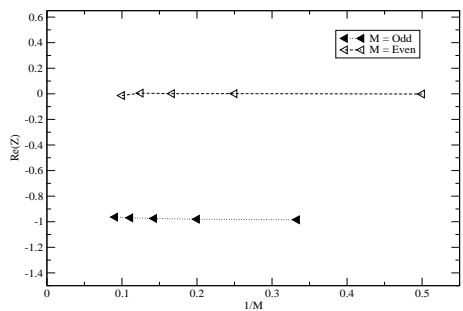}}
\caption{The VBS order parameter $z$ versus $1/M$, i.e. the
  concentration of $S^1=1$. The filled triangles are for the cases of
  $M=odd$ and empty ones are for $M=even$.}
\label{VBS}
\end{figure}
According to the LSM theorem, $z$ vanishes in the gapless
phase as system size $N\rightarrow \infty$. On the other hand, one
expects that $z$ varies in between $\pm 1$ but $z\neq 0$ in a
given gaped phase. In exact VBS states, $z=\pm 1$ \cite{Nakamura}.
Our calculations are plotted in Fig. \ref{VBS}. It is clear that $z
\approx -1$ for all cases of $M=odd$ to present the system located in
a VBS phase; while $z \approx 0$, it reveals the gapless energy
spectrum for all $M=even$ cases.

\begin{figure}[ht]
\epsfxsize=7cm\epsfysize=3.5cm \centerline{\epsffile{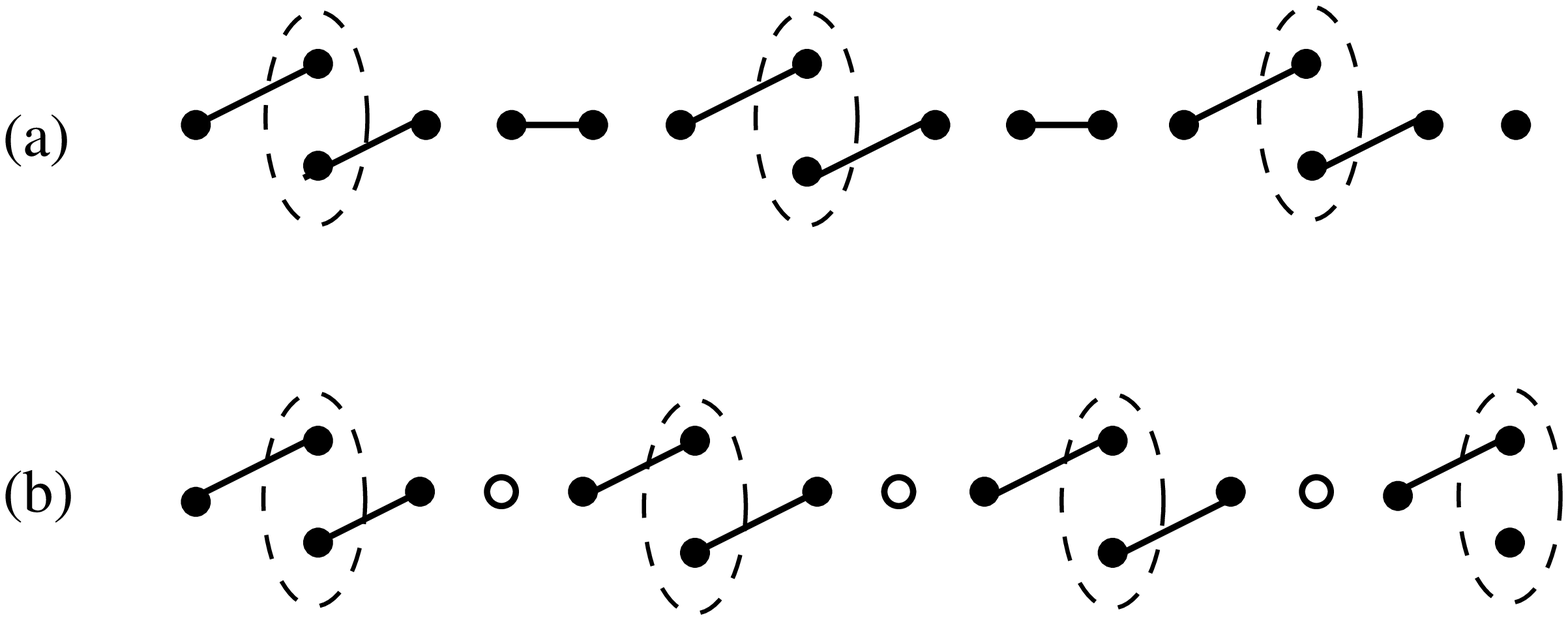}}
\caption{The illustrations of VBS picture. The dashed ellipses present
  the spin $S=1$, filled circles present the $S=1/2$ and empty circles
  present the active $S=1/2$ spins. (a) for $M=odd$ and (b) for
  $M=even$. The empty circles in (b) present the active spin-1/2.}
\label{illus_1}
\end{figure}

Especially, all these ground state phases can be understood under the
scenario of VBS picture. In VBS picture, each impurity $S^1=1$ can be
regarded as two spin-1/2 in a triplet state,  these two spin-1/2 can
form singlet with their nearest neighbor S=1/2 spin due to the antiferromagnetic
coupling. When $M=odd$, each unit have {\it even} number of S=1/2 host spins,
 so they can fall into singlets with their nearest neighbors
 including the two spin-1/2 of $S=1$ to induce the VBS order as
 seen in Fig. \ref{illus_1} (a). As a result, the system now shows a
 gaped energy spectrum.
But for $M=even$, {\it odd} number of spin-1/2 exist in a unit and
there will be an active spin which is not used to form singlet as
shown in Fig. \ref{illus_1}(b), thus there is no VBS order and the
system emerges no spin gap. Our results of the VBS order
parameter $z$ clearly verify this picture, $z \approx -1$ when
$M=odd$ and $z \approx 0$ when $M=even$  as shown in Fig.
\ref{VBS}. At last, we note that the VBS phase is stable with the
variation of $S^1=1$ concentration $\rho$ when $M=odd$.

\section{Discussion and Conclusion}
Our Monte Carlo study verifies that two brunches of different magnetic
behaviors emerge in cases of regular S=1 diluted S=1/2 host
chains. According Marshall theorem, the cases with
$M=even$ have the ferrimagnetic ground states which can be
specified by quantum number $S_{total}=|S_1-S_2|N/M$, so the
magnetization per site is finite and it decreases linearly as a
function of $\rho$ to ${\cal M}=0$, the case of the pure S=1/2
antiferromagnetic Heisenberg chain. This feature can be
easily observed from our results in Fig. \ref{mz} and Fig.
\ref{smz}. When $M$ is $odd$, the ground state is singlet with
$S_{total}=0$, thus the magnetization per site keeps zero and this
is a non-magnetic state. As observed in our simulations there is no
notable variations of the ground state magnetic properties in the
cases of $M=odd$.

To compare our ground state results of the model in this Letter
for $S^1=1$ and $S^2=1/2$ ({\it system I}) with the one we
studied previously \cite{Wang} when $S^1=1/2$ and $S^2=1$ ({\it
system II}),
 we collect the main points of the numerical calculations in Table I.
\begin{table}
\begin{center}
{\footnotesize
\begin{tabular}{|c|c|c|c|c|}
\hline
 &\multicolumn{2}{c|}{I. $S^1=1$, $S^2=1/2$} &\multicolumn{2}{c|}{II. $S^1=1/2$, $S^2=1$}\\
\hline
 & $M=odd$ & $M=even$ & $M=odd$ & $M=even$ \\
\hline
$S_{eff}$ & integer&half-integer&half-integer&half-integer\\
\hline
  $<$M$>$ & zero  & finite     & zero  & finite \\
\hline
$\chi_u$ & zero & large & small   & large \\
\hline
$<$M$_s$$>$&finite& finite & finite &finite \\
\hline
$\chi_s$  & small & large  & large  & large \\
\hline
$\Delta$&  gaped & gapless & gapless &gapless\\
\hline
$z$     & -1.0     & 0.0       & 0.0       & 0.0     \\
\hline
\end{tabular}
\caption{Comparison of ground state properties of two model, where
  $S_{eff}$ is effective spin in a unit, $<$M$>$ is magnetization, $<$M$_s$$>$ is staggered
  magnetization, $\chi_u$ is uniform susceptibility, $\chi_s$ is
  staggered susceptibility, $\Delta$ is energy gap and $z$ is VBS order parameter.}}
\end{center}
\end{table}
One can easily see that both systems behave with two kinds of different ground
state phases, magnetic or non-magnetic, respectively. If
$M=even$, the ground states are ferrimagnetic for both {\it system I} and
 {\it system II}, and their magnetizations and staggered
 magnetizations are all finite and decrease linearly with decreasing
 of impurity concentration. However, for the cases of $M=odd$, there
appears VBS order in {\it system I} which is gaped, but the
order is absent in {\it system II} where the spin arrangements can
not induce such order, so the gap is constantly closed. This
feature reveals that this topological order plays an important
role to the behavior of the energy gap in the mixed-spin system.
We believe that the fitted relation of $\Delta \approx 1.25
\sqrt{\rho}$, to denote the energy gap as function of $S^1=1$
concentration,  provides a good stuff to study how the
topological order affects the energy gap in the mixed-spin
systems.

In conclusion, we have studied the ground state and finite
temperature magnetic properties of the regular $S^1=1$ diluted
in $S^2=1/2$ antiferromagnetic chain. Our calculations show that there
exist different phases in the ground state as a function of $S^1$
concentration. When there is one $S^1$ impurity and $odd$ number
of host $S^2$ spins in a unit cell, the ground states are
ferrimagnetic and the system has a gapless energy spectrum. The
ferrimagnetism becomes weaker as the impurity concentration
reduced. While for one $S^1$ and $even$ number of $S^2$ in
one unit cell, the ground state is a VBS phase where there is a
gaped energy spectrum and the energy gap gradually approaches to
zero with decreasing the concentration $\rho$. An interesting
observation is that the behavior of the energy gap can be
numerically well fitted by $\Delta \approx 1.25 \sqrt{\rho}$.
Further analytical work, for exmaple using the mean-field theory
\cite{Wu}, is required to explain why such dependence of
the energy gap exist in VBS phases.
\section{Acknowledgment}
\vspace{-0.3cm}
The authors would like to thank Prof. Jianhui Dai for stimulating
discussions and comments. This work was supported in
part by the NNSF and SRFDP of China, and by the NSF of Zhejiang
province.


\begin{thebibliography}{99}
\bibitem{Sorensen} E. S. Sorensen and I. Affleck, Phys. Rev. B {\bf 51}, 16115(1995)
\bibitem{Ditusa} J.F. Ditusa, S.-W. Cheong, J.-H. Park, G. Aeppli, C. Broholm and C.T. Chen,
Phys. Rev. Lett. {\bf 73},1857(1994)
\bibitem{Wessel} S. Wessel and S. Haas, Phys. Rev. B {\bf 65},
  132402(2002)
\bibitem{Fukui} T. Fukui and N. Kawakami, Phys. Rev. B {\bf 55},
  R14709(1997); {\it ibid.} {\bf 56}, 8799(1997)
\bibitem{Takano} K. Takano, Phys. Rev. B {\bf 61}, 8863(2000).
\bibitem{Wu} C.-J. Wu, B. Chen, X. Dai, Y. Yu, and Z.-B. Su,
  Phys. Rev. B {\bf 60}, 1057(1999).
\bibitem{Yamamoto2} S. Yamamoto, T.Fukui, K. Maisinger and
  U. Schollwock, J. Phys.:Condens. Matter {\bf 10}, 11033(1997)
\bibitem{Pati} S.K. Pati, S. Ramasesha and D. Sen, Phys. Rev. B {\bf
  55}, 8894(1997); S.K. Pati, S. Ramasesha and D. Sen,
  J. Phys. Condens. Matter {\bf 9}, 8707(1997).
\bibitem{Tonegawa}T. Tonegawa, T. Hikihara, M. Kaburagi, T. Nishino,
  S. Miyashita, and H.-J. Mikeska, J. Phys. Soc. Jpn. {\bf 76},
  1000(1998).
\bibitem{Yamamoto} S. Yamamoto, J. Phys. Soc. Jpn. {\bf 64}, 4051(1995).
\bibitem{Kahn1} O. Kahn, Y. Pei and Y. Journaux, {\it Inorganic Materials},
  John Wiley $\and$ Sons Ltd., New York, 59-114 (1992).
\bibitem{Kahn2} O. Kahn, {\it Molecular Magnetism}, VCH, New York, 1993.
\bibitem{Verdaguer} M. Verdaguer, A. Gleizes, J. P. Renard and
  J. Seiden, Phys. Rev. B {\bf 29},5144-5155(1984).
\bibitem{Gleizes} A. Gleizes and M. Verdaguer, J. Am. Chem. Soc. {\bf
    106}, 3727(1984).
\bibitem{Pei} Y. Pei, M. Verdaguer, O. Kahn, J. Sletten and J.P. Renard,
  Inorg. Chem. {\bf 26},138(1987).
\bibitem{Hagiwara2} M. Hagiwara, K. Minami, Y. Narumi, K. Tatani and
  K. Kindo, J. Phys. Soc. Jpn. {\bf 67},2209-2211(1998).
\bibitem{Lieb}E.H. Lieb, T. Schultz and D.J. Mattis, Ann. Phys.
(New York){\bf 16}, 407(1961).
\bibitem{Wang} H.-W. Wang, Z.-X. Xu, H.-P. Ying and J. Zhang,
Intel. J. Mod. Phys. B {\bf 17}, 5951(2003)
\bibitem{Beard}B. B. Beard and U.-J. Wiese, Phys. Rev. Lett. {\bf 77},
  5130(1996)
\bibitem{Xu} Z.-X. Xu, J.-H. Dai, H.-P. Ying and B. Zheng,
  Phys. Rev. B {\bf 667}, 214426(2003); Z.-X. Xu, J. Zhang and
  H.-P. Ying, Commun. Theor. Phys. {\bf 40}, 623(2003).
\bibitem{Todo} S.Todo and K.Kato, Prog. Theor. Phys. Suppl. {\bf 138},
  535(2000).
\bibitem{Affleck} I. Affleck, T. Kennedy, E.H. Lieb and H. Tasaki,
  Phys. Rev. Lett. {\bf 59}, 799(1987)
\bibitem{Nakamura}M. Nakamura and S. Todo, cond-mat/0112377;
  Phys. Rev. Lett. {\bf 89}, 077204(2002).
\end{thebibliography}
\end{document}